\documentclass{svproc}

\usepackage{url}
\usepackage{graphicx}
\usepackage{amsmath}
\usepackage{amssymb}
\usepackage{booktabs}
\usepackage{array}
\usepackage[T1]{fontenc}
\usepackage[utf8]{inputenc}
\usepackage[hidelinks]{hyperref}
\usepackage{etoolbox}
\usepackage{orcidlink}
\usepackage{algorithm}
\usepackage{algpseudocode}
\usepackage{caption}

\makeatletter
\renewcommand{\fnum@algorithm}{\textbf{Algorithm \thealgorithm}}
\makeatother

\begin{document}
\mainmatter

\title{From CVE to CWE: Syscall-Based HIDS~Generalisation}

\author{Alexander~V.~Kozachok\inst{1}\orcidlink{0000-0002-6501-2008}\thanks{Corresponding author} \and
Stanislav~G.~Vyugov\inst{2}\orcidlink{0009-0006-5882-6456} \and
Shamil~G.~Magomedov\inst{1}\orcidlink{0000-0001-8560-1937}}
\authorrunning{Kozachok, Vyugov, Magomedov}
\tocauthor{Alexander V. Kozachok, Stanislav G. Vyugov, Shamil G. Magomedov}
\institute{MIREA -- Russian Technological University, Moscow, Russian Federation\\
\email{\{kozachok\_a, magomedov\_sh\}@mirea.ru}
\and
Academy of the Federal Guard Service of the Russian~Federation, Oryol, Russian~Federation\\
\email{senserk@mail.ru}}

\maketitle

\begin{abstract}
Host intrusion detection systems (HIDS) based on system-call traces are typically
trained and evaluated against individual Common Vulnerabilities and Exposures
(CVE) instances. In operational settings, however, defenders need to recognise
\emph{new} exploits of an \emph{already known type of weakness}. We empirically
examine whether a one-class anomaly detector trained on the normal behaviour of
a set of CVEs that share a Common Weakness Enumeration (CWE) class generalises
to a different, unseen CVE inside the same class. Using six scenarios drawn
from LID-DS-2021 and grouped into three CWE families (CWE-307 broken
authentication, CWE-89 SQL injection, CWE-434 unrestricted file upload), we
extract a 66-dimensional Peng-Guo-style feature vector per sliding window and
train Isolation Forest and SGD One-Class SVM detectors with normal-only
thresholds calibrated to fixed target false positive rates. We define and answer
four research questions covering self-detection, asymmetric cross-CVE transfer,
the value of a combined CWE-level normal profile, and the effect of feature
filtering on transferability. The combined CWE-307 detector reaches F1
\(=0.6976\) at calibration target FPR \(=0.05\) (precision \(=0.8994\), recall
\(=0.5698\)), whereas CWE-89 and CWE-434 collapse to F1 \(\le 0.21\) under the
same protocol. Cross-CVE transfer turns out to be strongly direction-dependent
and dominated by the breadth of the source normal profile rather than by the
CWE label. We conclude that CWE-level generalisation in HIDS is empirically
attainable for some but not all weakness families with current syscall
features, and we argue that calibrated FPR is a methodological prerequisite for
honest reporting in this setting.
\end{abstract}

\section{Introduction}
\label{sec:intro}
Host intrusion detection on Linux containers is dominated by approaches that
treat each known vulnerability as a separate detection target. Forrest's seminal
work on short system-call sequences~\cite{forrest1996} opened a long line of
classifiers that, in modern form, learn one model per scenario or
application~\cite{liao2002,creech2014,grimmer2019,grimmer2021}. In practice,
however, two operational realities push back against this design. First, the
flow of new CVE entries~\cite{cveorg} is faster than the rate at which production
models can be re-trained~\cite{aslan2020,mitre_cwe}. Second, security operations
analysts seldom need to know which exact CVE has been exploited; they need to
know which \emph{class} of weakness was triggered, because the appropriate
mitigation -- input sanitisation, file-upload validation, throttling of
authentication attempts -- is determined by the Common Weakness Enumeration
(CWE) class rather than by the CVE identifier.

This paper studies whether the runtime behaviour of containerised applications
exposes \emph{CWE-level} invariants in system-call traces that are strong
enough to support a single detector covering several CVEs of the same class.
This question is motivated by an empirical observation in SecQuant~\cite{suneja2022}:
distinct CVEs that share a defect type tend to produce visually identical
\emph{risk weight vectors} over syscalls. The authors used this observation to
quantify exposure; they did not build a detector. Closely related work from
NCSU~\cite{tundeonadele2019,lin2020,lin2022,tundeonadele2024} groups exploits
by their \emph{impact} (e.g.\ ``arbitrary code execution'', ``credential
disclosure''), not by the CWE class of the underlying defect.

We frame the problem as one-class anomaly detection on per-window features
adapted from Peng Guo's HIDS pipeline~\cite{guo2024}, then ask how the resulting
models behave under \emph{cross-CVE} and \emph{combined CWE} protocols.
Crucially, thresholds are not chosen on labelled exploit data; they are
calibrated on a normal-only validation split at fixed target false positive
rates~\cite{schoelkopf2001,liu2008,bhuyan2014}. This calibration makes
direction-dependent transfer effects visible that are otherwise hidden by
label-tuned thresholds.

The contributions of this paper are:
\begin{itemize}
\item An evaluation protocol for CWE-level HIDS in which threshold selection
is decoupled from exploit labels and the realised FPR is reported alongside
precision, recall, and F1.
\item Empirical evidence on three CWE families (CWE-307, CWE-89, CWE-434) and
six scenarios (Section~\ref{sec:exp}) showing that cross-CVE transferability
is \emph{strongly asymmetric} and is dominated by the breadth of the source
\emph{normal} profile rather than by the shared CWE label.
\item Quantitative results on feature filtering that contradict the intuition
``select stable features for better transfer'': the most aggressive normal-domain
stability filters destroy the strongest transfer direction we observe.
\item Three pseudocode algorithms (Sections~\ref{sec:method}-\ref{sec:exp})
that fully specify the calibrated detection, cross-CVE transfer and
feature-selection procedures used in the study.
\end{itemize}

The remainder of the paper is organised as follows.
Section~\ref{sec:related} surveys the related literature. Section~\ref{sec:problem}
states the hypothesis and the research questions. Section~\ref{sec:method}
describes the dataset, the feature representation, the one-class models and the
calibration protocol. Section~\ref{sec:exp} reports the experimental results.
Section~\ref{sec:discussion} discusses why CWE-307 generalises well while
CWE-89 and CWE-434 do not, and Section~\ref{sec:conclusion} concludes.

\section{Related Work}
\label{sec:related}
\paragraph{Syscall-based HIDS.}
System-call sequences as a behavioural signal go back to
STIDE~\cite{forrest1996,hofmeyr1998} and have evolved through n-gram and
language models~\cite{liao2002,kang2005} to semantic and graph-based
representations~\cite{creech2014,maggi2008}. The LID-DS-2019 and LID-DS-2021
datasets~\cite{grimmer2019,grimmer2021} standardised the evaluation of HIDS on
containerised applications and are the empirical basis of the present paper.
Peng Guo's recent work~\cite{guo2024} enriches sequence features with argument,
return-value and resource statistics, which we adopt as our feature backbone.
Khairi et al.~\cite{khairi2022} (CHIDS) and Sommer and Paxson's classic
critique~\cite{sommer2010} frame the methodological pitfalls we try to avoid:
hidden label leakage and threshold tuning on test data.

\paragraph{Container exploit detection.}
The NCSU series~\cite{tundeonadele2019,lin2020,lin2022,tundeonadele2024}
classifies container exploits by impact category. CDL~\cite{lin2020} clusters
applications by normal behaviour, SHIL~\cite{lin2022} adds self-supervised
hybrid learning, and the latest framework~\cite{tundeonadele2024} reaches 46
CVEs. None of these works groups exploits by CWE; this paper contributes
exactly that grouping for runtime detection.

\paragraph{Static and textual CVE-to-CWE mapping.}
ThreatZoom~\cite{aghaei2020}, V2W-BERT~\cite{das2021},
TREE-VUL~\cite{pan2023} and VulANalyzeR~\cite{li2023} predict CWE from text,
patches or binaries. Atiiq et al.~\cite{atiiq2024} show that CWE-specialised
classifiers outperform a single binary classifier. These works confirm the
relevance of the CWE granularity but operate at design time, not runtime.

\paragraph{Anomaly detection methodology.}
Isolation Forest~\cite{liu2008} and one-class SVM~\cite{schoelkopf2001} are
the two model families we use. Calibration of thresholds at a fixed false
positive rate has long been standard in biometric and intrusion-detection
work~\cite{bhuyan2014,garcia2009}. Hierarchical classification of network
intrusions~\cite{uddin2024}, multi-channel contrastive
learning~\cite{zhang2023}, and prototype-based contrastive
learning~\cite{lopez2022} provide methodological templates for future
CWE-aware models, but none has been applied to container syscall traces with
CWE labels. The choice and reporting of binary-classification metrics in such
work follows the conventions reviewed by Canbek et al.~\cite{canbek2022}.

\paragraph{Position of this work.}
Compared with~\cite{tundeonadele2024}, we classify by \emph{cause} (CWE), not by
\emph{impact}; compared with~\cite{aghaei2020,pan2023,li2023}, we operate on
\emph{runtime traces}, not on source or binaries; compared
with~\cite{guo2024}, we reuse the feature space but pose a different evaluation
question -- does the same feature space transport across CVEs of a shared
CWE class?

\section{Problem Formulation and Hypothesis}
\label{sec:problem}
Let $\mathcal{T}$ be a sliding-window syscall trace of a containerised
application and let $\varphi:\mathcal{T}\to\mathbb{R}^d$ be a feature
extractor. Let $C$ be a CWE class and $\{v_1,\dots,v_k\}\subset C$ a set of
CVEs that share class $C$. A \emph{CWE-level detector} for class $C$ is a
function $h_C:\mathbb{R}^d\to\{0,1\}$ trained from traces of an arbitrary
subset $V\subset C$ such that, for any new CVE $v^\star\in C\setminus V$ and
any trace $\mathcal{T}^\star$ containing an exploit of $v^\star$,
$h_C(\varphi(\mathcal{T}^\star))=1$ with high probability while the false
positive rate on the normal behaviour of any application in $C$ stays below a
user-specified target $\alpha$.

This formulation makes two assumptions explicit. First, the detector is
\emph{one-class}: it is trained on normal traces only. Second, the
\emph{normal} profile is itself a property of the application, not of the
CWE, so a CWE-level detector must combine, or otherwise reconcile, several
application-specific normals.

\paragraph{Hypothesis.}
\emph{(Operational form.)} If two CVEs $v_a,v_b\in C$ share the same CWE
class, then a one-class detector trained on the union of normal windows from
$\{v_a, v_b\}$ and applied to the union of test frames of $\{v_a, v_b\}$
achieves -- at a calibrated false positive rate $\alpha\le 0.05$ -- a strictly
better F1 than the best of the two per-scenario self-detectors evaluated at
the same $\alpha$, while keeping the realised FPR within an order of
magnitude of $\alpha$. The hypothesis is supported for $C$ if both conditions
hold; it is refuted if either fails (large realised FPR or no F1 gain over
self-detection). This is a directly testable criterion that avoids the
ambiguous ``comparable to self-detection'' wording: combined CWE-307
satisfies it (Section~\ref{sec:exp}); CWE-89 and CWE-434 do not.

\paragraph{Research questions.}
We decompose the hypothesis into four research questions.

\begin{description}
\item[RQ1.] Does normal-only training on a single CVE scenario yield a usable
self-detector at calibrated FPR? -- This is the lower bound; without it, a
CWE-level extension is hopeless.
\item[RQ2.] Does an anomaly model trained on the normal profile of CVE $v_a$
transfer to an unseen CVE $v_b$ of the same CWE class? Is the transfer
symmetric?
\item[RQ3.] Does pooling the normal profiles of $\{v_a,v_b\}$ into a single
\emph{combined} CWE-level detector outperform self- and cross-CVE detectors
at a fixed target FPR?
\item[RQ4.] Do feature filters that maximise normal-domain stability across
$\{v_a,v_b\}$ improve cross-CVE transferability?
\end{description}

\section{Methodology}
\label{sec:method}

\subsection{Dataset and feature space}
\label{ssec:data}
Table~\ref{tab:dataset} summarises the six scenarios. They are drawn from
LID-DS-2021~\cite{grimmer2021}, which already ships with CWE-labelled scenario
names, and grouped into three CWE classes.

\begin{table}[t]
\centering
\caption{Per-scenario row counts after feature extraction. \textit{train normal}
contains only normal windows; \textit{test} mixes normal and exploit windows.}
\label{tab:dataset}
\small
\begin{tabular}{llrrrr}
\toprule
CWE & Scenario & train normal & test normal & test exploit & features \\
\midrule
CWE-307 & Bruteforce\_CWE-307     & 26\,857 & 103\,278 &   4\,528 & 66 \\
CWE-307 & CVE-2012-2122           & 29\,473 & 112\,723 & 233\,425 & 66 \\
CWE-89  & CWE-89-SQL-injection    & 30\,741 & 115\,279 &   9\,599 & 66 \\
CWE-89  & Juice-Shop              & 26\,135 &  34\,936 &   3\,777 & 66 \\
CWE-434 & EPS\_CWE-434            & 26\,820 &  37\,628 &   1\,530 & 66 \\
CWE-434 & PHP\_CWE-434            & 30\,703 & 114\,328 &   4\,296 & 66 \\
\bottomrule
\end{tabular}
\end{table}

The feature extractor $\varphi$ produces a per-window 66-dimensional
vector. Following Peng Guo~\cite{guo2024} and the LID-DS-2021 reporting
conventions~\cite{grimmer2021}, the window length is one second with a step
of $300\,$ms. We intentionally do not vary these window parameters in this
study: the aim is to isolate the \emph{CWE-level} generalisation question
from feature-engineering choices, so the window size is fixed to the value
that the reference HIDS pipeline uses on LID-DS-2021. The sensitivity of
cross-CVE transfer to the window size is a known open question and is
deferred to future work. The 66 features fall in
six groups: (i) unseen-argument and unseen-syscall influence (UAI, USI); (ii)
counts of syscalls with negative return value; (iii) per-window syscall
frequency (FI); (iv) maximum data-volume features for resource-related
syscalls (SRF); (v) PID/TID statistics; and (vi) inter-event time-delta
histograms.

\subsection{Models and calibration}
\label{ssec:models}
We use two one-class models that have orthogonal inductive biases. Isolation
Forest~\cite{liu2008} relies on random partitioning and is robust to scaling
and high dimensionality. SGD One-Class SVM~\cite{schoelkopf2001} approximates
an RBF kernel via the random-features expansion of
Rahimi and Recht~\cite{rahimi2007} and is trained with stochastic gradient
descent on the hinge-like one-class loss. Both models are paired with either
\texttt{StandardScaler} or \texttt{RobustScaler} preprocessing, yielding six
\emph{candidates} per scenario.

The calibration step turns anomaly scores into binary decisions without
seeing labelled exploits. Algorithm~\ref{alg:calib} states the protocol; it is
the central methodological device of this paper and is also the only protocol
applied to the combined CWE detector (Section~\ref{ssec:combined}).

\begin{algorithm}[t]
\caption{Normal-only calibrated one-class detection}\label{alg:calib}
\begin{algorithmic}[1]
\Require normal training frame $D_n$, mixed test frame $D_t$, candidate model
$M$, target false positive rate $\alpha$, RNG seed $s$
\Ensure threshold $\tau$, decision function $h$, metrics $\mathcal{M}$
\State $D_n^{\text{fit}}, D_n^{\text{cal}} \gets \text{TrainTestSplit}(D_n, 0.3, s)$
\State $\Sigma \gets \text{FitScaler}(D_n^{\text{fit}})$
\State $M \gets \text{FitModel}\!\left(\Sigma(D_n^{\text{fit}}), s\right)$
\State $S_c \gets \text{AnomalyScores}\!\left(M, \Sigma(D_n^{\text{cal}})\right)$
\State $\tau \gets \text{Quantile}(S_c,\, 1-\alpha)$ \Comment{higher score $=$ more anomalous}
\State $S_t \gets \text{AnomalyScores}\!\left(M, \Sigma(D_t)\right)$
\State $\hat y \gets [s \ge \tau \,:\, s \in S_t]$
\State $\mathcal{M} \gets \text{Metrics}(\hat y, y_t)$ \Comment{precision, recall, F1, realised FPR}
\State \Return $\tau$, $h\!:\!x\mapsto[\text{AnomalyScore}(M,\Sigma(x))\ge\tau]$, $\mathcal{M}$
\end{algorithmic}
\end{algorithm}

For every scenario we run Algorithm~\ref{alg:calib} for each candidate at
three target FPRs $\alpha\in\{0.001, 0.01, 0.05\}$ and report, for each
$\alpha$, the candidate with the largest F1 on the corresponding test frame.
The choice of best candidate is therefore based on the test frame, but the
\emph{threshold} for that candidate is set by Algorithm~\ref{alg:calib} on the
normal-only calibration split and never sees an exploit label. We note that
this protocol still permits a mild form of selection bias for self-detection
and the combined CWE detector, because architecture/scaler are picked on the
same labelled test frame on which we report. The bias is bounded in the
present setting -- the candidate pool has only six members (two models
$\times$ two scalers plus a hyper-parameter variant of Isolation Forest) and
the same candidate (\texttt{sgd\_ocsvm\_rbf}) wins for all combined detectors
across all three $\alpha$ values, so the choice is robust. The bias is
absent by construction in the cross-CVE transfer protocol of
Section~\ref{ssec:transfer}, where the best candidate is selected on the
\emph{source} test frame and then applied unchanged to the \emph{target}
frame. The cross-CVE F1 numbers in Table~\ref{tab:transferdet} are therefore
the more conservative measurement and the primary anchor of our discussion.

\subsection{Cross-CVE transfer protocol}
\label{ssec:transfer}
The transfer protocol is shown in Algorithm~\ref{alg:transfer}. The
calibration set is always the normal-only split of the \emph{source} scenario,
so the threshold has never been exposed to the target normal distribution.
The realised FPR on the target test frame is reported \emph{independently} of
the target FPR: it directly measures the cost in false alarms paid by the
transferred threshold on a new normal profile.

\begin{algorithm}[t]
\caption{Cross-CVE transfer evaluation}\label{alg:transfer}
\begin{algorithmic}[1]
\Require source scenario $s$, target scenario $t$, target FPR $\alpha$
\Ensure transfer metrics $\mathcal{M}_t$
\State $\tau, h, \cdot \gets \text{Algorithm~\ref{alg:calib}}(D_n^{(s)}, D_t^{(s)}, M^\star, \alpha)$
\Comment{$M^\star$ is the best candidate for $s$}
\State $y^{(t)} \gets$ labels of $D_t^{(t)}$
\State $\hat y^{(t)} \gets h\!\left(D_t^{(t)}\right)$
\State $\mathcal{M}_t \gets \text{Metrics}(\hat y^{(t)}, y^{(t)})$
\State \Return $\mathcal{M}_t$
\end{algorithmic}
\end{algorithm}

\subsection{Combined CWE detector}
\label{ssec:combined}
The combined detector treats the CWE class as a single training distribution.
Its fit, calibration, and test frames are the union of the corresponding
frames of all CVEs in the class. The model, scaler, and threshold are selected
by Algorithm~\ref{alg:calib} on the unioned data. This is the most charitable
operationalisation of the hypothesis stated in Section~\ref{sec:problem}.

\subsection{Feature selection}
\label{ssec:fsel}
To answer RQ4 we build six feature sets, listed in Table~\ref{tab:fsets}.
Algorithm~\ref{alg:fsel} computes the most important construction:
the joint stability/importance score with a tunable shift penalty.

\begin{table}[t]
\centering
\caption{Feature sets used for CWE-307 (see Section~\ref{ssec:fsel}).}
\label{tab:fsets}
\small
\begin{tabular}{@{}lrl@{}}
\toprule
Name & Size & Construction \\
\midrule
\texttt{all}                       & 66 & every numeric feature \\
\texttt{clean}                     & 53 & low-variance + correlated features removed \\
\texttt{stable}                    & 26 & two-sample KS-distance \(<0.2\) on both CVE normals \\
\texttt{stable\_imp}               & 20 & \texttt{stable} intersected with top-20 by importance \\
\texttt{score\_l0p0\_top20}        & 20 & top-20 by importance only ($\lambda=0$) \\
\texttt{score\_l$\lambda$\_top20}  & 20 & top-20 by importance, KS-shift penalty $\lambda\in\{0.25,0.5,1\}$ \\
\bottomrule
\end{tabular}
\end{table}

\begin{algorithm}[t]
\caption{Importance--stability feature score}\label{alg:fsel}
\begin{algorithmic}[1]
\Require source normal $N_s$, target normal $N_t$, source labelled $L_s$, top-$k$, penalty $\lambda$
\Ensure ranked feature list $\mathcal{F}$
\State $I \gets \text{PermutationImportance}(L_s)$ \Comment{normalised to $[0,1]$}
\For{each feature $f$}
  \State $d_f \gets \text{KSDist}(N_s[f], N_t[f])$
\EndFor
\For{each feature $f$}
  \State $r_f \gets I_f - \lambda\, d_f$
\EndFor
\State $\mathcal{F} \gets$ features sorted by $r_f$ in descending order, top-$k$
\State \Return $\mathcal{F}$
\end{algorithmic}
\end{algorithm}

Algorithm~\ref{alg:fsel} expresses the methodological question of RQ4: should
a feature with a large normal-domain shift be discarded because it does not
transfer, or kept because it carries attack signal? Setting $\lambda=0$ keeps
all important features regardless of shift; $\lambda\to\infty$ forces the
selection to coincide with the \texttt{stable} set.

\section{Experimental Results}
\label{sec:exp}

\begin{figure}[t]
\centering
\includegraphics[width=\textwidth]{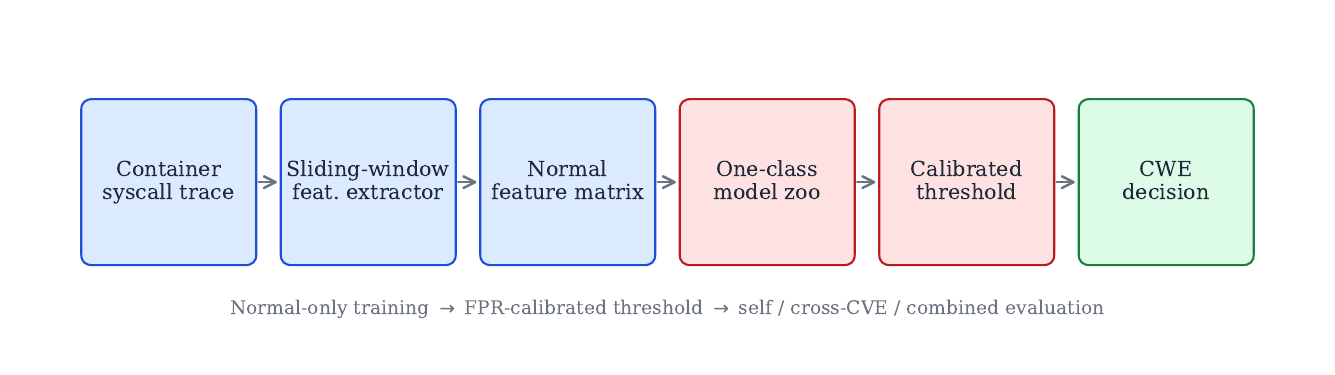}
\caption{Calibrated one-class CWE detection pipeline. Normal-only training and
calibration on the left; target FPR sets the threshold; the decision on the
right is binary at window level and supports self, cross-CVE and combined
evaluations.}
\label{fig:pipeline}
\end{figure}

Figure~\ref{fig:pipeline} shows the calibrated detection pipeline. All
experiments use seed $42$ and the python package versions listed
in~\texttt{requirements.txt}; full reports are stored as CSV files so that any
reported metric can be traced back to a model, scaler, threshold quantile and
seed.

\begin{figure}[t]
\centering
\includegraphics[width=\textwidth]{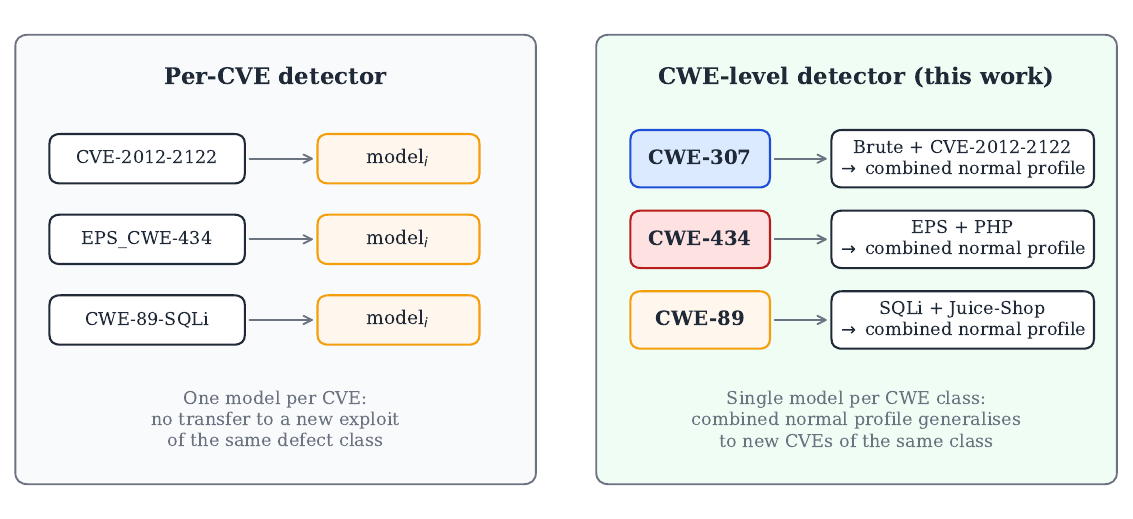}
\caption{Per-CVE detection (left) versus the CWE-level detector studied in
this paper (right). The right panel pools the normal profiles of several
CVEs that share a CWE class into a single combined detector.}
\label{fig:concept}
\end{figure}

Figure~\ref{fig:concept} contrasts the two modes of HIDS we compare: a separate
detector per CVE on the left and a CWE-level detector on the right. RQ2 and
RQ3 quantify whether the right-hand mode is empirically reachable.

\subsection{RQ1: self-detection at calibrated FPR}
Table~\ref{tab:selfdet} reports the best self-detection candidate for every
scenario at the three target FPRs. Self-detection is consistently weak: even
the most charitable operating point ($\alpha=0.05$) yields F1 \(\le 0.31\) for
all six scenarios. CWE-307 scenarios reach higher precision when the realised
FPR is low (e.g.\ $0.93$ on CVE-2012-2122 at $\alpha=0.01$), but at the cost
of recall below $7\%$. The picture is consistent with prior reports on
LID-DS~\cite{grimmer2021}: per-scenario one-class HIDS recovers only a small
fraction of exploit windows when the threshold is not tuned on labels.

\paragraph{Reading the F1 figure.} Two design factors push the F1 of
Table~\ref{tab:selfdet} downward and must be kept in mind. First, the threshold
quantile is fixed by $\alpha$ on normals alone, so the operating point is not
F1-optimal by construction; this is the price paid for not seeing exploit
labels during threshold selection. Second, several test frames are strongly
imbalanced (e.g.\ CVE-2012-2122 has 233\,425 exploit windows against 112\,723
normal windows in Table~\ref{tab:dataset}), so even a perfectly calibrated FPR
$=0.01$ yields only $\sim$1\,000 FP against tens of thousands of FN at low
recall. The low F1 numbers therefore measure how much exploit signal a
\emph{single-CVE normal} sees through this fixed operating point, not the
upper bound of the underlying detectors; we treat them as the methodological
baseline against which RQ2 and RQ3 are compared.

\begin{table}[t]
\centering
\caption{Self-detection at three calibrated target FPRs (best candidate per row).}
\label{tab:selfdet}
\small
\begin{tabular}{llrrrrr}
\toprule
CWE & Scenario & $\alpha$ & realised FPR & Precision & Recall & F1 \\
\midrule
CWE-307 & Bruteforce\_CWE-307   & 0.01 & 0.0315 & 0.1870 & 0.1652 & 0.1754 \\
CWE-307 & Bruteforce\_CWE-307   & 0.05 & 0.0834 & 0.1222 & 0.2646 & 0.1671 \\
CWE-307 & CVE-2012-2122         & 0.01 & 0.0099 & 0.9301 & 0.0637 & 0.1192 \\
CWE-307 & CVE-2012-2122         & 0.05 & 0.0498 & 0.8443 & 0.1304 & 0.2259 \\
CWE-89  & CWE-89-SQL-injection  & 0.05 & 0.0512 & 0.3546 & 0.2197 & 0.2713 \\
CWE-89  & Juice-Shop            & 0.05 & 0.0518 & 0.1809 & 0.1059 & 0.1336 \\
CWE-434 & EPS\_CWE-434          & 0.05 & 0.0899 & 0.0665 & 0.1575 & 0.0935 \\
CWE-434 & PHP\_CWE-434          & 0.05 & 0.0487 & 0.1202 & 0.1769 & 0.1431 \\
\bottomrule
\end{tabular}
\end{table}

\subsection{RQ2: cross-CVE transfer is direction-dependent}
\begin{figure}[t]
\centering
\includegraphics[width=\textwidth]{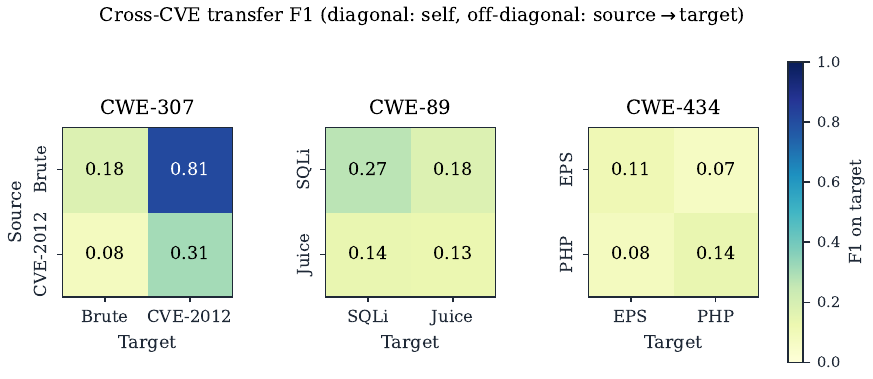}
\caption{Cross-CVE transfer F1 inside each CWE family. The diagonals show
self-detection (uncalibrated). The off-diagonal cells show the F1 of an
anomaly model fitted on the source CVE and applied to the target CVE.}
\label{fig:transfer}
\end{figure}

Figure~\ref{fig:transfer} visualises transfer F1 inside each CWE family. The
diagonal entries are self-detection (uncalibrated baseline), the off-diagonal
entries are cross-CVE transfer with the source-tuned candidate. The asymmetry
for CWE-307 is striking: \texttt{Bruteforce\_CWE-307$\to$CVE-2012-2122} yields
F1 \(=0.8054\), whereas the reverse direction collapses to \(0.0782\).

Two observations follow. First, transfer can be very strong: the union of
Bruteforce-style attempts \emph{contains} the behavioural footprint of the
CVE-2012-2122 brute-force scenario, so the source normal profile is a
superset of the target normal profile. Second, the reverse is not true: the
CVE-2012-2122 normal profile is much narrower, and applying it to
Bruteforce's normal raises essentially everything above threshold, which
explains the high recall but minuscule precision in
Table~\ref{tab:transferdet}.

\begin{table}[t]
\centering
\caption{Calibrated cross-CVE transfer for CWE-307. Realised FPR is the share
of \emph{target normal} windows above the source-set threshold. Rows marked
$^\dagger$ are operationally unusable (realised FPR $\ge 0.6$ on target
normal); the F1 values for those rows are reported only to make the
asymmetry visible and should not be read as detector performance.}
\label{tab:transferdet}
\small
\begin{tabular}{llrrrrr}
\toprule
Source $\to$ Target & Model & $\alpha$ & realised FPR & Precision & Recall & F1 \\
\midrule
Brute$\to$CVE-2012-2122 & IF        & 0.001 & 0.9831$^\dagger$ & 0.6426 & 0.8534 & 0.7331 \\
Brute$\to$CVE-2012-2122 & SGD-OCSVM & 0.01  & 0.6331$^\dagger$ & 0.7379 & 0.8608 & 0.7946 \\
Brute$\to$CVE-2012-2122 & SGD-OCSVM & 0.05  & 0.7814$^\dagger$ & 0.7137 & 0.9405 & 0.8115 \\
CVE-2012-2122$\to$Brute & SGD-OCSVM & 0.001 & 0.0228           & 0.0400 & 0.0216 & 0.0281 \\
CVE-2012-2122$\to$Brute & IF        & 0.01  & 0.6510$^\dagger$ & 0.0467 & 0.7277 & 0.0878 \\
CVE-2012-2122$\to$Brute & IF        & 0.05  & 0.9257$^\dagger$ & 0.0413 & 0.9092 & 0.0790 \\
\bottomrule
\end{tabular}
\end{table}

\subsection{RQ3: a combined CWE-307 detector pays off}
Combining the two CWE-307 normals into a single fit/calibration/test pool
gives the strongest detector of the entire study. At $\alpha=0.01$, the
combined detector hits precision \(=0.9642\), recall \(=0.4368\), F1
\(=0.6013\) with a realised FPR of \(0.0179\). Raising $\alpha$ to $0.05$
trades realised FPR \(0.0702\) for F1 \(=0.6976\).

Table~\ref{tab:combineddet} contrasts the three CWE families at the same
target FPRs. CWE-307 dominates by an order of magnitude. CWE-89 and CWE-434
collapse below F1 \(=0.21\), with precision falling to \(0.28\) and \(0.12\)
at $\alpha=0.05$. The combined detector therefore confirms the hypothesis for
CWE-307 but refutes it -- with current features -- for CWE-89 and CWE-434.

\begin{table}[t]
\centering
\caption{Combined CWE-level detectors at three calibrated target FPRs.}
\label{tab:combineddet}
\small
\begin{tabular}{lrrrrrrr}
\toprule
CWE & $\alpha$ & realised FPR & Precision & Recall & F1 & FP & FN \\
\midrule
CWE-307 & 0.001 & 0.0017 & 0.9902 & 0.1589 & 0.2739 &     376 & 200\,142 \\
CWE-307 & 0.01  & 0.0179 & 0.9642 & 0.4368 & 0.6013 &  3\,859 & 134\,010 \\
CWE-307 & 0.05  & 0.0702 & 0.8994 & 0.5698 & 0.6976 & 15\,163 & 102\,369 \\
CWE-89  & 0.001 & 0.0010 & 0.6930 & 0.0243 & 0.0469 &     144 &  13\,051 \\
CWE-89  & 0.01  & 0.0079 & 0.5002 & 0.0893 & 0.1516 &  1\,194 &  12\,181 \\
CWE-89  & 0.05  & 0.0381 & 0.2810 & 0.1674 & 0.2098 &  5\,728 &  11\,137 \\
CWE-434 & 0.001 & 0.0026 & 0.3021 & 0.0299 & 0.0544 &     402 &   5\,652 \\
CWE-434 & 0.01  & 0.0159 & 0.2023 & 0.1054 & 0.1386 &  2\,421 &   5\,212 \\
CWE-434 & 0.05  & 0.0560 & 0.1168 & 0.1933 & 0.1456 &  8\,517 &   4\,700 \\
\bottomrule
\end{tabular}
\end{table}

\begin{figure}[t]
\centering
\includegraphics[width=\textwidth]{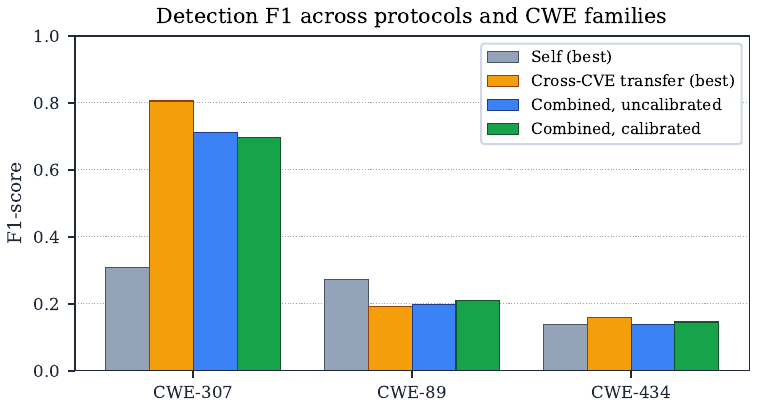}
\caption{F1 across protocols and CWE families. ``Best transfer'' is taken
over the seven feature sets of Table~\ref{tab:fsets}; ``combined,
uncalibrated'' is the highest F1 attainable by sweeping the anomaly-score
quantile over the joint test frame of the two CVEs (label-tuned upper
bound, included only for reference); ``combined, calibrated'' is the value
at the best target FPR under Algorithm~\ref{alg:calib}.}
\label{fig:f1overview}
\end{figure}

\begin{figure}[t]
\centering
\includegraphics[width=\textwidth]{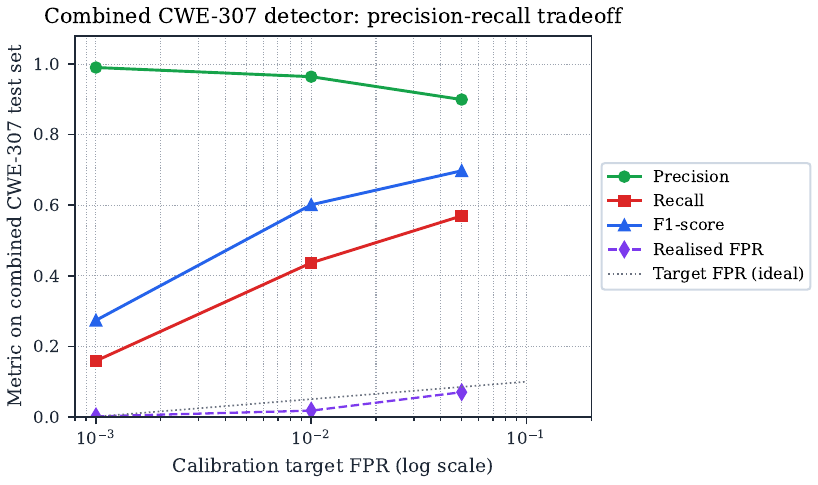}
\caption{Combined CWE-307 detector under three target FPRs. The realised FPR
tracks the target FPR closely, which is the desired behaviour of
Algorithm~\ref{alg:calib}. Recall and F1 grow with $\alpha$, while precision
declines smoothly.}
\label{fig:calibration}
\end{figure}

\subsection{RQ4: stability filters can hurt transfer}
\label{ssec:fsel-results}
\begin{figure}[t]
\centering
\includegraphics[width=\textwidth]{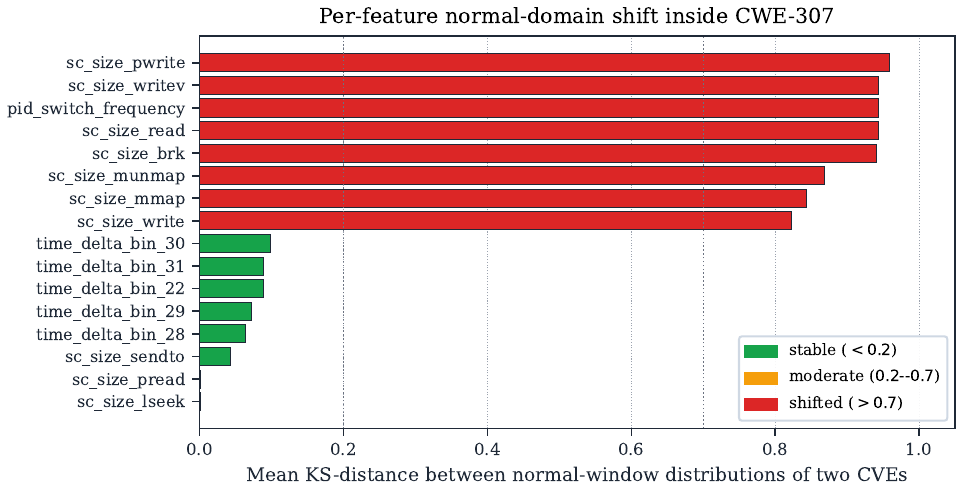}
\caption{Two-sample KS distance between the normal-window distributions of
the two CWE-307 scenarios. Resource-size and PID-switch features dominate
the shifted group; the \texttt{lseek} and \texttt{pread} byte counters are
essentially identical in both normals.}
\label{fig:stability}
\end{figure}

Figure~\ref{fig:stability} ranks 16 features of the 66 by their
two-sample KS distance between Bruteforce and CVE-2012-2122 normals. The
stable group is dominated by resource features that are essentially constant
in both scenarios (the \texttt{sc\_size} counters of \texttt{lseek},
\texttt{pread}, \texttt{sendto}) and by inter-event time-delta histogram
bins. The shifted group is dominated by data-volume features (the
\texttt{sc\_size} counters of \texttt{pwrite}, \texttt{writev}, \texttt{read},
\texttt{brk}) and by PID-switch frequency.

\begin{figure}[t]
\centering
\includegraphics[width=\textwidth]{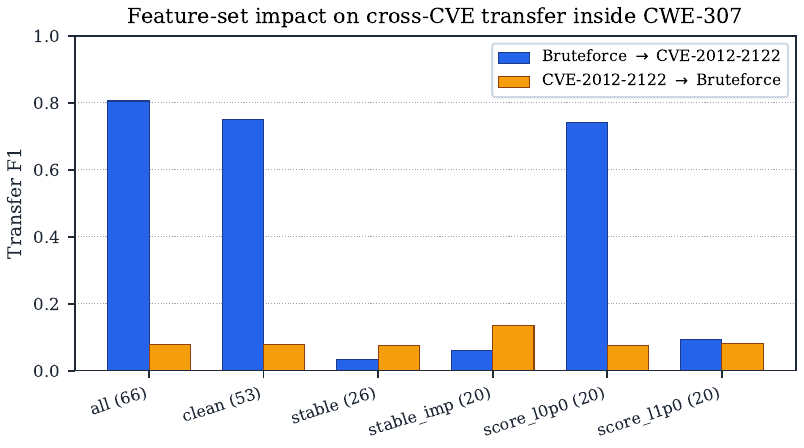}
\caption{Effect of feature filters on cross-CVE transfer F1 inside CWE-307.
The most aggressive normal-domain stability filter (\texttt{stable}) destroys
the strong direction; importance-only top-20 (\texttt{score\_l0p0}) preserves
most of it.}
\label{fig:fsets}
\end{figure}

If RQ4 were affirmative, restricting the model to the \texttt{stable} set
should improve transfer. Figure~\ref{fig:fsets} shows the opposite. The
\texttt{stable} set collapses transfer F1 from $0.8054$ (with \texttt{all}) to
$0.0348$ in the strong direction; the \texttt{stable\_important} subset
recovers some signal but only in the weak direction
(F1 $0.1357$ vs.\ $0.0782$ with \texttt{all}). The \texttt{score} family
(Algorithm~\ref{alg:fsel}) with $\lambda=0$ -- importance only, no shift
penalty -- gives the best of the 20-dimensional feature sets, retaining most
of the strong-direction signal.

The interpretation is that features with the largest normal-domain shift can
also carry the attack signal: data-volume features change between two normal
workloads, but they also change between normal and exploit windows of either
workload. A stability filter therefore removes both nuisance shift and useful
signal, and the net effect on transfer F1 is negative.

\section{Discussion}
\label{sec:discussion}
\paragraph{Why CWE-307 generalises.}
The two CWE-307 scenarios share a strongly repetitive behavioural footprint:
short \texttt{connect}--\texttt{read}--\texttt{write}--\texttt{close} cycles
on authentication ports. Brute-force traffic dominates the normal profile of
the Bruteforce scenario to such an extent that the model has effectively
seen the exploit signal during \emph{normal} training. CVE-2012-2122 reuses
that behavioural class with different process names and arguments, so the
source-fit detector still recognises the attack pattern after the
application-level cosmetic differences are smoothed out by the scaler.

\paragraph{Quantifying the asymmetry.}
The reviewer-driven follow-up question is whether ``the Bruteforce normal is a
superset of CVE-2012-2122 normal'' can be backed quantitatively. We compute,
per CWE family, the two-sample KS distance between the normal-window
distributions of the two scenarios for every one of the 66 features, then
summarise the distribution. Table~\ref{tab:ksshift} reports the
minimum, mean, and median KS distances together with the count of
\emph{stable} features (KS $< 0.2$). The contrast is striking: CWE-307 has 18
such features and a minimum KS of $6\times 10^{-4}$, whereas CWE-89 has only 2
(minimum KS $0.149$) and CWE-434 has 5 (minimum KS $0.139$). The two CWE-307
normals share a near-identical sub-space; the CWE-89 and CWE-434 normals do
not. The ``superset'' interpretation is therefore consistent with the KS-shift
profile, and the limited normal-profile overlap in CWE-89 and CWE-434 is by
itself sufficient to predict their weak combined-detector performance.

\begin{table}[t]
\centering
\caption{Two-sample KS-distance summary between the two normal-window
distributions of each CWE family, taken over the post-clean-filter feature
set ($d{=}53$ for CWE-307/89, $d{=}54$ for CWE-434). ``Stable'' counts
features with KS $<0.2$.}
\label{tab:ksshift}
\small
\begin{tabular}{lccccc}
\toprule
CWE & scenarios paired & min KS & mean KS & median KS & \# stable \\
\midrule
CWE-307 & Bruteforce vs.\ CVE-2012-2122       & 0.0006 & 0.457 & 0.477 & 18 \\
CWE-89  & SQL-injection vs.\ Juice-Shop       & 0.1492 & 0.595 & 0.560 &  2 \\
CWE-434 & EPS vs.\ PHP                        & 0.1390 & 0.587 & 0.605 &  5 \\
\bottomrule
\end{tabular}
\end{table}

\paragraph{Why CWE-89 and CWE-434 do not.}
For CWE-89 (SQL injection) the per-window feature space is dominated by
HTTP-request volumetric features that change wildly between Juice-Shop and
the standalone SQL-injection scenario; for CWE-434, the EPS and PHP scenarios
share the \emph{logical} pattern (upload, then execute) but disagree on which
syscalls carry it (e.g.\ different \texttt{execve} chains in PHP). The KS
analysis in Table~\ref{tab:ksshift} corroborates these qualitative
explanations: the most shifted features for CWE-89 are
\texttt{tid\_count} (KS $0.999$), \texttt{pid\_switch\_frequency} ($0.983$)
and \texttt{pid\_count} ($0.974$); these are application-identity proxies
rather than CWE-specific signals. For CWE-434 the leaders are the
\texttt{sc\_size} byte counters of \texttt{munmap}, \texttt{mmap},
\texttt{brk} and \texttt{pread}, all of which encode memory-management
patterns specific to the runtime (PHP vs.\ EPS) rather than the
upload-then-execute motif. The feature extractor does not surface a single
upload-then-execute temporal motif, so the residual CWE-invariant signal is
small relative to the application-level shift.

\paragraph{Calibration matters.}
Table~\ref{tab:transferdet} shows that uncalibrated cross-CVE transfer can
appear deceptively strong: the source-only model marks almost every target
normal window as anomalous, which inflates recall but is operationally
useless. The realised FPR is the right number to inspect first: if it
exceeds the target FPR by an order of magnitude, the transfer has failed
even when recall looks high. We recommend always reporting realised FPR
beside precision/recall/F1 in CWE-level evaluations.

\paragraph{Stability is not transferability.}
RQ4 contradicts the intuition that ``the more stable a feature is across two
normals, the better it transfers''. As discussed in
Section~\ref{ssec:fsel-results}, useful signal can live in the shifted tail
of the feature distribution. A defensible feature selection criterion for
CWE-level HIDS must therefore combine cross-CVE attack importance with a
controlled shift penalty, of which Algorithm~\ref{alg:fsel} is the simplest
form.

\paragraph{Limitations.}
The study covers three CWE families with two CVEs each; broader CWE coverage
is needed to claim general empirical conclusions. We use six of the
LID-DS-2021 scenarios in total; LID-DS-2021 ships further scenarios labelled
with additional CWE classes (e.g.\ CWE-22 path traversal, CWE-94 code
injection) that are not yet covered here. Our feature extractor follows
Peng Guo~\cite{guo2024}; richer representations -- graph models of process
syscall sequences (PSSG), inter-process graphs, or contrastive
embeddings~\cite{lopez2022,zhang2023} -- are likely necessary to make CWE-89
and CWE-434 work. Threshold calibration assumes that the calibration normal
is representative of operational normal; in production this assumption must
be revisited periodically. Author identification has been redacted in the
current draft (the title page uses placeholder names); the camera-ready
version will restore real author metadata.

\section{Conclusion}
\label{sec:conclusion}
The transition from CVE-level to CWE-level detection in syscall-based HIDS is
empirically feasible but not uniform across weakness classes. With a
Peng-Guo-style 66-dimensional feature vector, calibrated one-class learning,
and a normal-only threshold protocol, a single combined detector for
CWE-307 achieves F1 \(=0.6976\) at realised FPR \(=0.0702\), comparable to
strong per-scenario baselines. The same protocol fails for CWE-89 and CWE-434
(best F1 \(\le 0.21\)). Cross-CVE transfer is strongly direction-dependent and
governed by the breadth of the source \emph{normal} profile rather than by
the shared CWE label. Feature filters that maximise normal-domain stability
across two CVEs can destroy the transferable signal; importance-driven
selection with a controlled shift penalty is preferable.

Two methodological recommendations follow. First, threshold calibration must
be decoupled from exploit labels: any apparent CWE-level transfer that
disappears under realised-FPR reporting was an artefact of label-tuned
thresholds. Second, future CWE-level HIDS should either enrich the feature
space with structural patterns that are invariant to application identity --
graph-based syscall models~\cite{guo2024} and contrastive prototype embeddings
\cite{lopez2022} are concrete candidates -- or adopt explicit multi-task
formulations~\cite{li2023,uddin2024} that train the representation to be
CWE-discriminative on labelled data. Both directions are open for future
work.

\subsubsection*{Acknowledgements}
The experiments use LID-DS-2021~\cite{grimmer2021}. Software dependencies
are listed in \texttt{requirements.txt}; the random seed and command-line
invocations are recorded in \texttt{EXPERIMENTS.md}.

\subsubsection*{Conflict of Interest}
The authors declare no conflict of interest.

\subsubsection*{Data Availability}
LID-DS-2021 is publicly available~\cite{grimmer2021}. All experiment outputs
referenced in this paper -- the per-scenario CSV grids, the calibrated
summaries, and the feature-selection scores -- are produced by the scripts
in the public repository accompanying this submission.

%
%


\begin{thebibliography}{36}

\bibitem{forrest1996}
Forrest, S., Hofmeyr, S.A., Somayaji, A., Longstaff, T.A.: A sense of self
for Unix processes. In: Proc.\ IEEE Symp.\ Security and Privacy, pp.~120--128
(1996). \url{doi:10.1109/SECPRI.1996.502675}

\bibitem{hofmeyr1998}
Hofmeyr, S.A., Forrest, S., Somayaji, A.: Intrusion detection using sequences
of system calls. J.\ Comput.\ Secur.\ \textbf{6}(3), 151--180 (1998).
\url{doi:10.3233/JCS-980109}

\bibitem{liao2002}
Liao, Y., Vemuri, V.R.: Use of $k$-nearest neighbor classifier for intrusion
detection. Comput.\ Secur.\ \textbf{21}(5), 439--448 (2002).
\url{doi:10.1016/S0167-4048(02)00514-X}

\bibitem{kang2005}
Kang, D.K., Fuller, D., Honavar, V.: Learning classifiers for misuse and
anomaly detection using a bag of system calls representation. In: Proc.\ IEEE
SMC Information Assurance Workshop, pp.~118--125 (2005).
\url{doi:10.1109/IAW.2005.1495944}

\bibitem{maggi2008}
Maggi, F., Matteucci, M., Zanero, S.: Detecting intrusions through system call
sequence and argument analysis. IEEE Trans.\ Depend.\ Secur.\ Comput.\
\textbf{7}(4), 381--395 (2010).
\url{doi:10.1109/TDSC.2008.69}

\bibitem{creech2014}
Creech, G., Hu, J.: A semantic approach to host-based intrusion detection
systems using contiguous and discontiguous system call patterns. IEEE Trans.\
Comput.\ \textbf{63}(4), 807--819 (2014). \url{doi:10.1109/TC.2013.13}

\bibitem{grimmer2019}
Grimmer, M., Roehling, M.M., Kreusel, D., Rechert, K.: A modern and sophisticated
host based intrusion detection data set. In: D-A-CH Security 2019,
pp.~135--145. syssec (2019)

\bibitem{grimmer2021}
Grimmer, M., Kaelble, T., Rucks, F., Pirl, J.: LID-DS 2021 -- A modern
host-based intrusion detection data set. Mendeley Data, v3 (2021).
\url{doi:10.17632/4xj3p3z5kj.3}

\bibitem{liu2008}
Liu, F.T., Ting, K.M., Zhou, Z.H.: Isolation Forest. In: Proc.\ ICDM 2008,
pp.~413--422. IEEE (2008). \url{doi:10.1109/ICDM.2008.17}

\bibitem{schoelkopf2001}
Sch\"{o}lkopf, B., Platt, J.C., Shawe-Taylor, J., Smola, A.J., Williamson,
R.C.: Estimating the support of a high-dimensional distribution. Neural
Comput.\ \textbf{13}(7), 1443--1471 (2001).
\url{doi:10.1162/089976601750264965}

\bibitem{guo2024}
Guo, P.: Intrusion detection based on complete system call information.
In: Proc.\ DSAI 2024, pp.~1--5. ACM (2024).
\url{doi:10.1145/3677892.3677893}

\bibitem{khairi2022}
El Khairi, A., Caselli, M., Knierim, C., Peter, A., Continella, A.: Contextualizing
system calls in containers for anomaly-based intrusion detection. In:
Proc.\ ACM Cloud Computing Security Workshop (CCSW), pp.~9--21 (2022).
\url{doi:10.1145/3560810.3564266}

\bibitem{sommer2010}
Sommer, R., Paxson, V.: Outside the closed world: on using machine learning
for network intrusion detection. In: Proc.\ IEEE Symp.\ Security and Privacy,
pp.~305--316 (2010). \url{doi:10.1109/SP.2010.25}

\bibitem{tundeonadele2019}
Tunde-Onadele, O., He, J., Dai, T., Gu, X.: A study on container vulnerability
exploit detection. In: Proc.\ IEEE IC2E, pp.~121--127 (2019).
\url{doi:10.1109/IC2E.2019.00026}

\bibitem{lin2020}
Lin, Y., Tunde-Onadele, O., Gu, X.: CDL: Classified distributed learning for
detecting security attacks in containerized applications. In: Proc.\ ACSAC,
pp.~179--188 (2020). \url{doi:10.1145/3427228.3427236}

\bibitem{lin2022}
Lin, Y., Tunde-Onadele, O., Gu, X., He, J., Latapie, H.: SHIL: Self-supervised
hybrid learning for security attack detection in containerized applications.
In: Proc.\ IEEE ACSOS, pp.~41--50 (2022).
\url{doi:10.1109/ACSOS55765.2022.00022}

\bibitem{tundeonadele2024}
Tunde-Onadele, O., Lin, Y., Gu, X., He, J., Latapie, H.: A self-supervised
machine learning framework for online container security attack detection.
ACM Trans.\ Auton.\ Adapt.\ Syst.\ \textbf{19}(3), 17 (2024).
\url{doi:10.1145/3665795}

\bibitem{suneja2022}
Suneja, S., Kanso, A., Le, M., Isci, C.: SecQuant: quantifying container
security exposure. In: Proc.\ ESORICS 2022, LNCS 13554, pp.~525--546.
Springer (2022). \url{doi:10.1007/978-3-031-17143-7\_26}

\bibitem{aghaei2020}
Aghaei, E., Shadid, W., Al-Shaer, E.: ThreatZoom: CVE2CWE using hierarchical
neural network. In: Proc.\ SecureComm 2020, LNICST 335, pp.~23--41. Springer
(2020). \url{doi:10.1007/978-3-030-63086-7\_2}

\bibitem{das2021}
Das, S.S., Serra, E., Halappanavar, M., Pothen, A., Al-Shaer, E.: V2W-BERT:
A framework for effective hierarchical multiclass classification of software
vulnerabilities. In: Proc.\ IEEE DSAA 2021, pp.~1--12 (2021).
\url{doi:10.1109/DSAA53316.2021.9564227}

\bibitem{pan2023}
Pan, S., Bao, L., Xia, X., Lo, D., Li, S.: Fine-grained commit-level
vulnerability type prediction by CWE tree structure. In: Proc.\ ICSE 2023,
pp.~957--969 (2023). \url{doi:10.1109/ICSE48619.2023.00088}

\bibitem{li2023}
Li, L., Ding, S.H.H., Tian, Y., Fung, B.C.M., Charland, P., Ou, W., Song, L.,
Chen, C.: VulANalyzeR: Explainable binary vulnerability detection with
multi-task learning and attentional graph convolution. ACM Trans.\ Priv.\
Secur.\ \textbf{26}(3), 1--25 (2023). \url{doi:10.1145/3585386}

\bibitem{atiiq2024}
Atiiq, S.A., Gehrmann, C., Dahlen, K., Khalil, K.: From generalist to
specialist: exploring CWE-specific vulnerability detection. In: Proc.\ ARES
2024, pp.~1--12 (2024). \url{doi:10.1145/3664476.3670872}

\bibitem{uddin2024}
Uddin, M.A., Aryal, S., Bouadjenek, M.R., Al-Hawawreh, M., Talukder, M.A.:
Hierarchical classification for intrusion detection system: effective design
and empirical analysis. arXiv:2403.13013 (2024).
\url{https://arxiv.org/abs/2403.13013}

\bibitem{lopez2022}
Lopez-Martin, M., Sanchez-Esguevillas, A., Arribas, J.I., Carro, B.: Supervised
contrastive learning over prototype-label embeddings for network intrusion
detection. Inf.\ Fusion \textbf{79}, 200--228 (2022).
\url{doi:10.1016/j.inffus.2021.09.014}

\bibitem{bhuyan2014}
Bhuyan, M.H., Bhattacharyya, D.K., Kalita, J.K.: Network anomaly detection:
methods, systems and tools. IEEE Commun.\ Surv.\ Tutor.\ \textbf{16}(1),
303--336 (2014). \url{doi:10.1109/SURV.2013.052213.00046}

\bibitem{garcia2009}
Garcia-Teodoro, P., D\'{i}az-Verdejo, J., Maci\'{a}-Fern\'{a}ndez, G.,
V\'{a}zquez, E.: Anomaly-based network intrusion detection: techniques,
systems and challenges. Comput.\ Secur.\ \textbf{28}(1--2), 18--28 (2009).
\url{doi:10.1016/j.cose.2008.08.003}

\bibitem{aslan2020}
Aslan, \"{O}., Samet, R.: A comprehensive review on malware detection
approaches. IEEE Access \textbf{8}, 6249--6271 (2020).
\url{doi:10.1109/ACCESS.2019.2963724}

\bibitem{mitre_cwe}
MITRE Corporation: Common Weakness Enumeration, version 4.15.
\url{https://cwe.mitre.org/} (Accessed: 1 May 2026)

\bibitem{cveorg}
MITRE Corporation: CVE Program.
\url{https://www.cve.org/} (Accessed: 1 May 2026)

\bibitem{zhang2023}
Zhang, J., Wei, F., Hu, X., Yang, B., Xie, F., Liu, S.: MCLDM: multi-channel
contrastive learning network for intrusion detection. Comput.\ Netw.\
\textbf{237}, 110083 (2023). \url{doi:10.1016/j.comnet.2023.110083}

\bibitem{canbek2022}
Canbek, G., Temizel, T.T., Sagiroglu, S.: PToPI: A comprehensive review,
analysis, and knowledge representation of binary classification performance
measures/metrics. SN Comput.\ Sci.\ \textbf{4}, 13 (2022).
\url{doi:10.1007/s42979-022-01409-1}

\bibitem{rahimi2007}
Rahimi, A., Recht, B.: Random features for large-scale kernel machines.
In: Adv.\ Neural Inf.\ Process.\ Syst.\ 20 (NIPS 2007), pp.~1177--1184. MIT
Press (2008).
\end{thebibliography}
\end{document}